\documentclass{elsart}
\usepackage{amssymb}
\usepackage{graphicx}
\usepackage{dcolumn}
\usepackage{bm}
\newcommand{\wo}{\omega_{\mathrm{0}}}
\newcommand{\fracc}[2]{\frac{\displaystyle #1}{\displaystyle #2}}
\renewcommand{\d}{\mathrm{d}}
\newcommand{\E}{\tilde{E}}
\newcommand{\V}{\tilde{V}}
\renewcommand{\v}{\tilde{v}}
\newcommand{\nnu}{\tilde{\nu}}
\newcommand{\hh}{\hat{h}}

\newcommand{\tW}{\tilde W}
\newcommand{\WW}{\hat W}
\newcommand{\CC}{\hat C}
\newcommand{\GG}{\hat G}
\renewcommand{\a}{\tilde a}
\renewcommand{\b}{\tilde b}

\begin{document}

\newcommand{\cm}{cm$^{-1}$}
 \newcommand{\A}{\AA$^{-1}$}
 \newcommand{\muev}{$\mu$eV}
  \newcommand{\mueV}{$\mu$eV}
\begin{frontmatter}

\title{Crystal structures and proton dynamics
 in potassium and cesium hydrogen bistrifluoroacetate
  salts with strong symmetric hydrogen bonds}

\author{Alain Cousson}
 \ead{cousson@llb.saclay.cea.fr}
\address{Laboratoire L\'{e}on Brillouin (CEA-CNRS),
 C.E. Saclay, 91191 Gif-sur-Yvette cedex, France}

\author{Juan F R Archilla}
 \address{Group of Nonlinear Physics,
 Departamento de F\'{\i}sica Aplicada I, ETSI Inform\'atica.
 Avda Reina Mercedes s/n, 41012-Sevilla, Spain}
 \ead{archilla@us.es}

\author{John Tomkinson}
\address{Rutherford Appleton Laboratory, Chilton,
 Didcot, OX11 0QX, United Kingdom}
\ead{J.Tomkinson@rl.ac.uk}

\author{Fran\c{c}ois Fillaux}\footnote{Corresponding author \\ homepage http://ulysse.glvt-cnrs.fr/ladir/pagefillaux.htm}
\ead{fillaux@glvt-cnrs.fr}
\address{LADIR-CNRS, UMR 7075 Universit\'{e}
 Pierre et Marie Curie, 2 rue Henry Dunant, 94320 Thiais, France}

\date{\today}

\newpage

\begin{abstract}
The crystal structures of potassium and cesium bistrifluoroacetates
were determined at room temperature and at 20 K and 14 K, respectively,
with the single crystal neutron diffraction technique.
The crystals belong to the $I2/a$ and $A2/a$ monoclinic space groups,
respectively, and there is no visible phase transition. For both crystals,
the trifluoroacetate entities form dimers linked by very short hydrogen
bonds lying across a centre of inversion. Any proton disorder or double
minimum potential can be rejected. The inelastic neutron scattering
spectral profiles in the OH stretching region between 500 and 1000 \cm\
previously published [Fillaux and Tomkinson, Chem. Phys. 158 (1991) 113]
are reanalyzed. The best fitting potential has the major characteristics
already reported for potassium hydrogen maleate [Fillaux \textit{et al.}
Chem. Phys. 244 (1999) 387]. It is composed of a narrow well containing
the ground state and a shallow upper part corresponding to dissociation
of the hydrogen bond.
\end{abstract}

\begin{keyword}
Single crystals neutron diffraction, inelastic
neutron scattering, hydrogen bond, proton dynamics
\end{keyword}
\end{frontmatter}

\newpage

\section*{\label{sec:01}Introduction}
In most of hydrogen bonds AH$\cdots$B, it is widely accepted that the
potential energy for the motion of the hydrogen atom has two
minima \cite{anb,pml,VL,jano,mm}. For symmetric systems
A$\cdots$H$\cdots$A, the two wells can be equivalent and proton
tunnelling may occur. For the shortest hydrogen bonds the distance
between the two minima decreases and the potential barrier between
them may disappear \cite{Novak}. There remains a single well
at the centre and such symmetric structures
can be regarded as ``intermediate-states'' for proton transfer in
chemical reactions or biological processes \cite{cle,ck}.
Our motivation for the studies of the title compounds reported below
is to shed light on the potential function experienced by protons
involved in strong symmetric hydrogen bonds.

A classic example of a symmetric intramolecular hydrogen bond is the
mono-anion of maleic acid in potassium hydrogen maleate,
KH(OOC-CH=CH-COO). This extremely short O$\cdots$H$\cdots$O bond, with
length R$_{\textrm{o}\cdots\textrm{o}}$ = 2.437 \AA\ and with the
proton located at the centre, has been studied with different
techniques: X-ray diffraction
\cite{dc,dar}; infrared and Raman \cite{CDO,nsb,aokh,zmi,ibrb} and
inelastic neutron scattering (INS) \cite{tbhw,htegt,tph}. More
recently, the shape of the potential function for the proton motion along the hydrogen bond has
been determined from neutron diffraction and INS spectra of single
crystals at low temperature \cite{FLTCP}. This potential is
composed of a sharp well at the centre of the hydrogen bond and
two secondary minima at $\approx 1000$ \cm\ above the central
minimum. These secondary minima are located at $\approx \pm 0.8$
\AA\ from the centre. Upon the assumption that these secondary minima
occur because the hydrogen bond geometry is no longer stable after
excitation of the $\nu$ OH mode above 500 \cm, it could be concluded
that whilst the ground state wave function is ``hydrogen-bonding''
the excited vibrational states are ``hydrogen-nonbonding''. The
potential shape is a snapshot of the proton motion during the
dissociation of the hydrogen bond upon excitation of the OH
stretching mode. The apparent paradox of a modest dissociation
energy threshold observed in one of the ``strongest'' known
hydrogen bonds was ascribed to partial compensation of the energy
of formation of the hydrogen bond by a strain energy arising from
the planarity of the maleate ring. This potential function was criticized
by Wilson \textit{et al.} \cite{WTM} on the basis of plane-wave
DFT calculations. However, such calculations performed over a very limited
range of proton displacement, corresponding to potential energy far below
the ground state, are not conclusive \cite{FCT}.

So far, potassium hydrogen maleate is the only reported example we are aware of this
unforeseen potential shape and whether this system is unique or if similar
potentials also apply to other strong symmetric hydrogen bonds is an open
question.

By contrast, a totally different potential shape has been proposed for
potassium and cesium hydrogen bistrifluoroacetates,
KH(CF$_3$COO)$_2$ and \\ CsH(CF$_3$COO)$_2$, which are also classic
examples of strong and symmetric hydrogen bonds at room temperature \cite{Novak} with
lengths R$_{\textrm{o}\cdots\textrm{o}}$ = 2.435 and 2.38 \AA,
respectively \cite{GS}. The vibrational spectra of these salts
have been thoroughly investigated
\cite{HON,FT}. Upon comparison of infrared, Raman and INS spectra, the
vibrational modes for protons were assigned at $\approx 800$ \cm\
($\nu$ OH), 1300 \cm\ ($\gamma$ OH) and 1600 \cm\ ($\delta$ OH). For both
systems, the $\nu$ OH band profiles observed with INS and in the
infrared are almost identical. They show
complex multi-component profiles with main components at $\approx 700$,
800 and 880 \cm\ and weaker components in the range 500-1000 \cm.
These profiles were regarded as mainly due to Evans-type
\cite{Evans1,Evans2} or Fano-type \cite{Fano} interactions with
internal modes of the CF$_3$COO entities. For the Cs derivative, a
very sharp INS band at $\approx 90$ \cm\ was
tentatively assigned to proton tunnelling along the OH stretching
coordinate and an asymmetric double minimum potential was
proposed, quite at variance to the case of potassium hydrogen
maleate.

However, with hindsight, such a double well is rather
unlikely for very short hydrogen bonds because the minima located
at $\approx \pm 0.5$ \AA\ off centre (see Fig.~7 in Ref.~\cite{FT})
are quite incompatible with the R$_{\textrm{o}\cdots\textrm{o}}$
length. The O---H distance less than 0.7
\AA\ is quite unrealistic. In addition, there is no evidence
of $\nu$ OH tunnelling transition
for the K derivative, whereas band profiles, crystal
structures and hydrogen bond lengths are very similar.

The localization of the proton in an asymmetric double well is
also in conflict with the known crystal structures determined with
X-ray diffraction \cite{GS}: hydrogen bonds are centrosymmetric
for the two salts at room temperature. However, crystal structures
at low temperature are unknown and a symmetry breakdown
cannot be excluded. In order to determine the proton location we
have used the single crystal neutron diffraction technique. The
crystal structures of the two salts at 20 K or 14 K and at 298 K
(room temperature) are presented below. In both systems, hydrogen
bonds are symmetrical at any temperature, with protons clearly
located at the centre. It is shown that potential functions
largely inspired by that of potassium hydrogen maleate, though
significantly  different, account for the OH stretching INS band
profiles.

\section{\label{sec:02}Crystal structures}

Single crystals were obtained by slow recrystallization from
aqueous solutions. These very hygroscopic colorless crystals were
handled under dry atmosphere. For neutron diffraction
measurements, approximately cubic samples ($3\times 3\times 3$
mm$^3$) were cut from large crystals and tested at room
temperature. Samples were loaded into aluminum containers
that were then mounted in a cryostat. For measurements at low
temperature, they were cooled down with a flow of helium vapor.

Measurements (see Table \ref{tab:1}) were carried out on a Stoe
four-circle diffractometer 5C2 at the Orph\'{e}e reactor
(Laboratoire L\'{e}on-Brillouin) \cite{LLB}. The incident neutron
wavelength was $\lambda = 0.8305$ \AA. Absorption corrections were
ignored. Data analysis was carried out with the computer
package CRYSTALS \cite{BCCPW}

\begin{figure}[p]
\includegraphics[width=\textwidth]{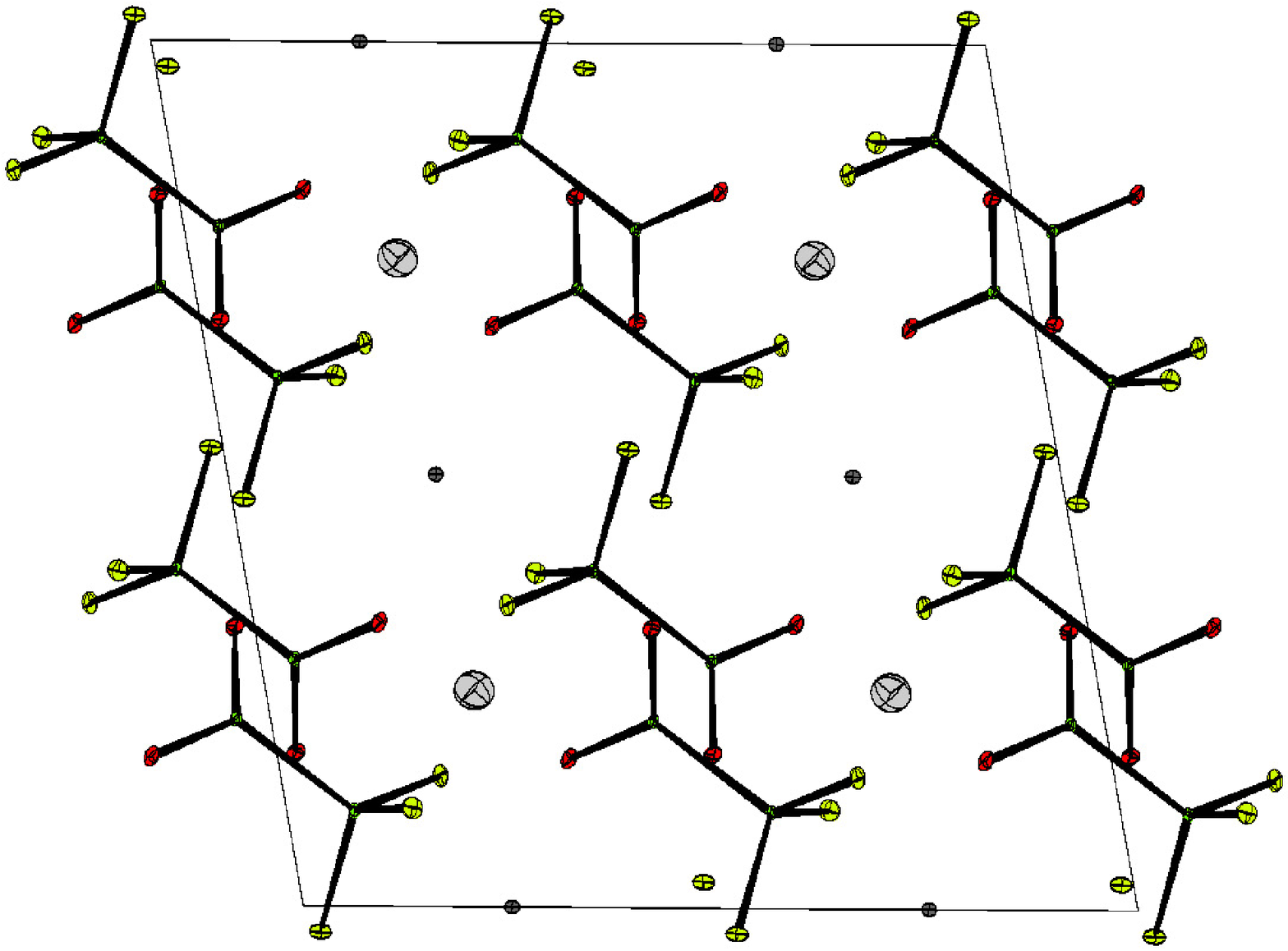}
\caption{\label{fig:01} Projection of the crystal structure of
potassium hydrogen bistrifluoroacetate at 20 K onto the $(a,c)$
plane and thermal ellipsoids for atoms.}
\end{figure}

\begin{figure}[p]
\includegraphics[width=\textwidth]{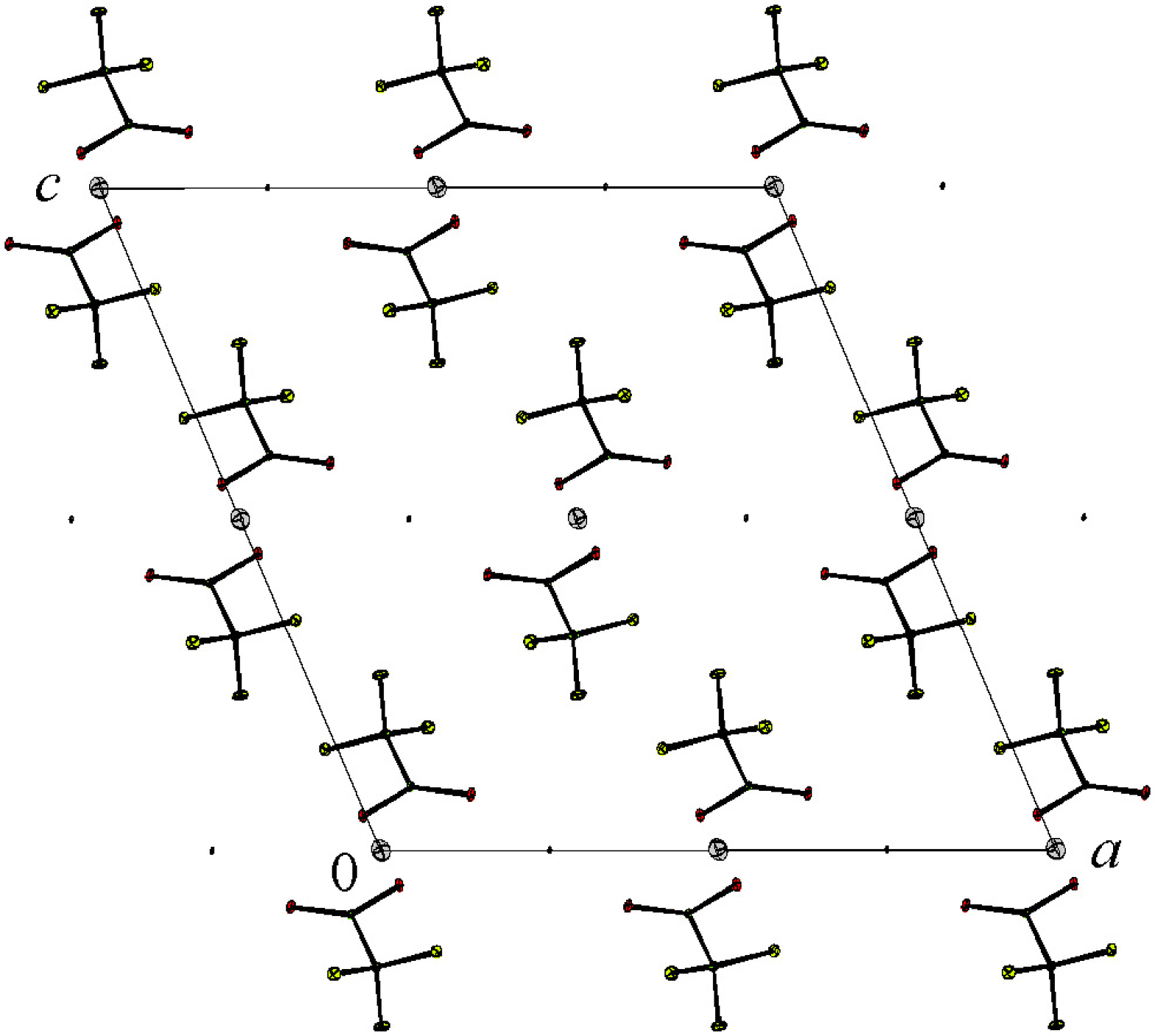}
\caption{\label{fig:02} Projection of the crystal structure of
cesium hydrogen bistrifluoroacetate at 14 K onto the $(a,c)$
plane and thermal ellipsoids for atoms.}
\end{figure}

\begin{figure}[p]
\includegraphics[width=\textwidth]{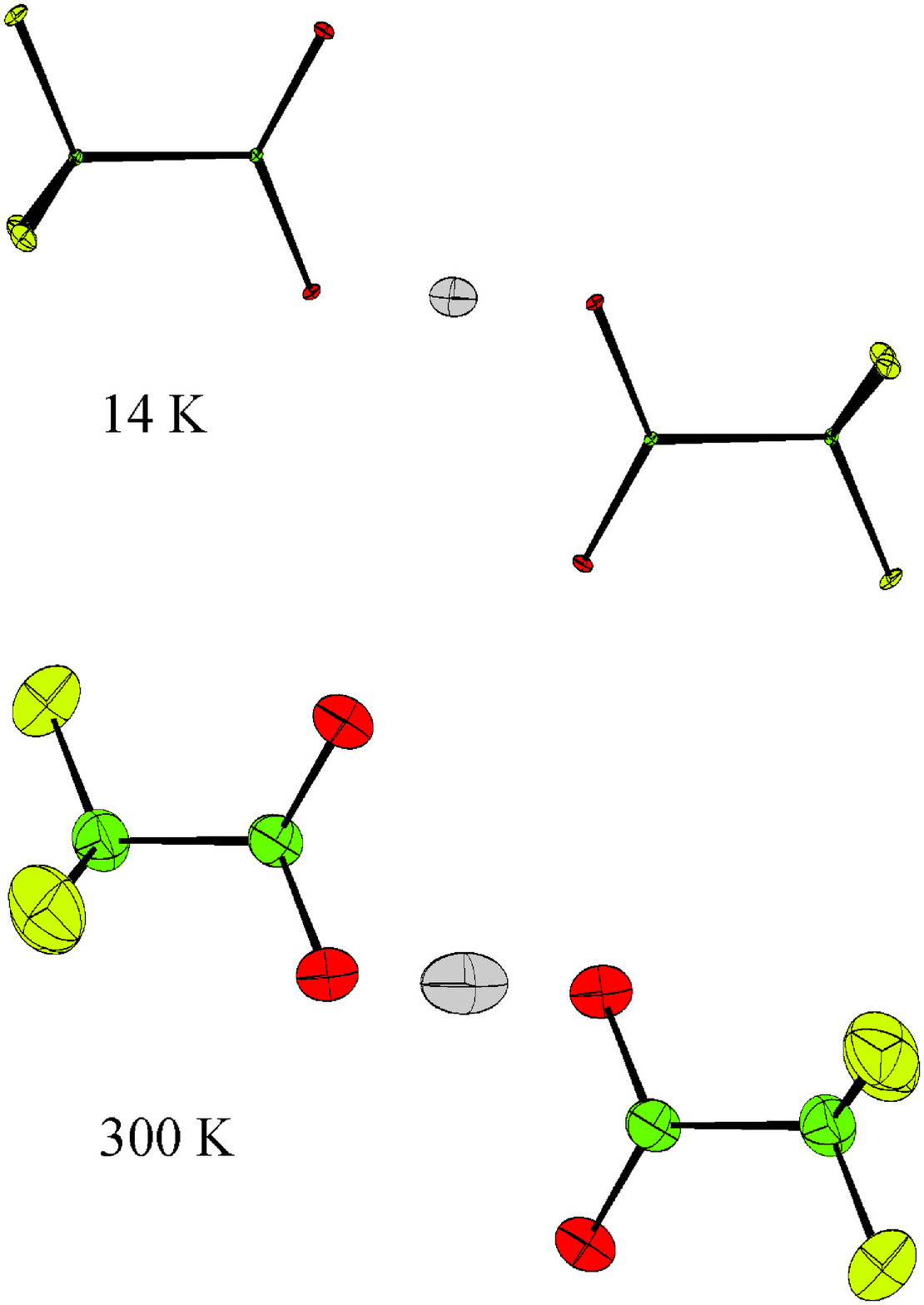}
\caption{\label{fig:03} Comparison of the hydrogen bistrifluoroacetate
entities and thermal ellipsoids for atoms of the potassium salt at 20 and 300 K.
Right: projection onto the mean-plane of the carboxylic entities. Left: view
along the hydrogen bond direction.}
\end{figure}

\begin{figure}[p]
\includegraphics[width=\textwidth]{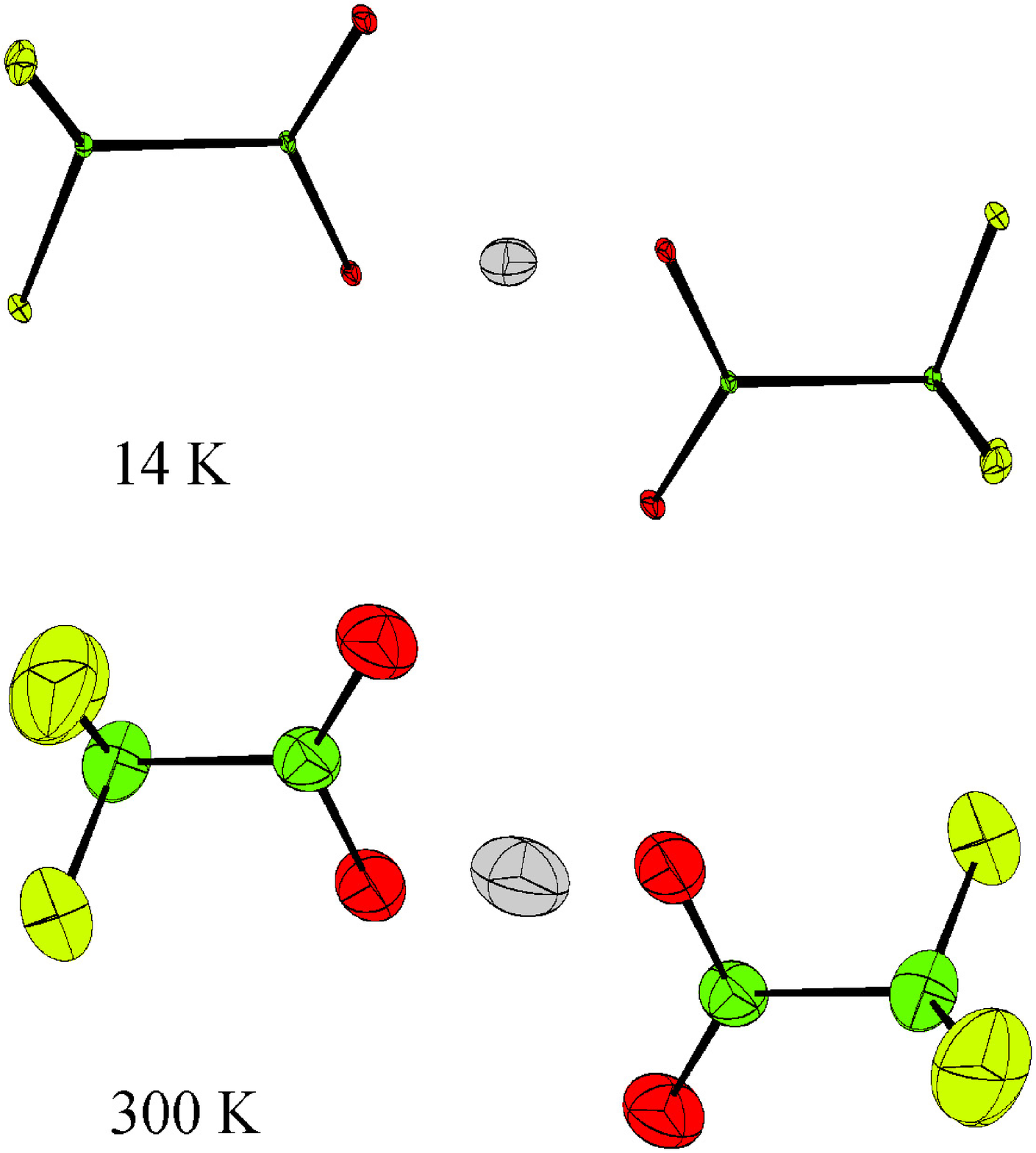}
\caption{\label{fig:04} Comparison of the hydrogen bistrifluoroacetate
 entities and thermal ellipsoids for atoms of the cesium salt at 14 and 300 K.
 Right: projection onto the mean-plane of the carboxylic entities. Left: view
along the hydrogen bond direction.}
\end{figure}

Crystal structures at any temperature are
similar to those previously determined with X-ray diffraction
\cite{GS}. The potassium and cesium salts belong to the $I2/a$ and
$A2/a$ monoclinic space groups, respectively, both with four entities in the
unit cells (see Figs.~\ref{fig:01} and \ref{fig:02}). The two acetate
residues are crystallographically equivalent for both salts. They are
linked by short hydrogen bonds lying across a centre of inversion
(see Figs. \ref{fig:03} and \ref{fig:04}). Hydrogen bond lengths are
identical at low temperature: R$_{\textrm{o}\cdots\textrm{o}}$ = 2.436(3)
and 2.436 (4) \AA, for the K and Cs derivatives, respectively (see Tables
\ref{tab:3} and \ref{tab:6}). At room
temperature, they are slightly different: 2.432(3) and 2.444(2) \AA,
respectively. However, the difference is tiny compared to the
thermal factors for O atoms ($\approx 0.05$ \AA$^2$, see Tables \ref{tab:4}
and \ref{tab:7}).

At any temperature there is no evidence for splitting of the proton probability
density that could be attributed to disorder over different
sites or to a double well potential. At low temperature, the temperature
factors for oxygen atoms are much smaller than those for protons. As the latter
compare to the mean square amplitudes of
displacements for proton oscillators with averaged frequencies of $\approx
1000$ \cm\ (see Tables \ref{tab:4} and \ref{tab:7}), they are primarily
due to proton dynamics in a single well. At room temperature,
the temperature factors increase substantially (see Figs. \ref{fig:03} and
\ref{fig:04}). For protons these factors can be regarded as combinations of
the proton contribution, virtually unchanged compared to that at low temperature, with those of
the oxygen atoms (those labelled O(2) in Tables \ref{tab:4} and \ref{tab:7}), which increase dramatically.
Therefore, even at room temperature, the temperature factors are not
compatible with any realistic double well potential, with minima
separated by more than 0.5 \AA\ \cite{FT}.

Finally, trifluoroacetate entities have
different conformations for the two salts (see Figs. \ref{fig:03} and
\ref{fig:04}). A C---F bond is \textit{trans} with
respect to the CO(H) bond for the K salt. One of the C---F bonds is
virtually perpendicular to the carboxylic plane for the Cs derivative.

\section{\label{sec:03}INS band profiles and proton dynamics}

The (CF$_3$COO$\cdots$H$\cdots$OOCCF$_3$)$^-$ entities occupying
$Ci$ sites, the OH stretching and bending modes are only infrared
active ($Au+Bu$) \cite{HON}. The stretching mode gives a broad
continuum of intensity centered at $\approx 800$ \cm, whereas
bending modes are extremely weak and barely visible. This
assignment scheme is corroborated by the correlation between $\nu$
OH frequencies and hydrogen bond lengths, R$_{\mathrm{O}\cdots
\mathrm{O}}$ \cite{Novak}. It is also consistent with the OH
stretching at $\approx 600$ \cm\ for the maleate salt
\cite{FLTCP}.

In the crystal structures, the shortest distances between protons
are 4.34 and 6.72 \A\ for K and Cs salts, respectively. These
protons are further isolated by the stacking of CF$_3$ entities
and K$^+$ or Cs$^+$ ions. Dynamical coupling between protons is
negligible. The great similarity of the $\nu$ OH profiles observed
in the infrared and with INS (free of symmetry related selection
rule) confirms that crystal field splitting is not resolved
\cite{FT}. Moreover, the profiles are almost identical for the two
salts (see below), though they have different crystal structures.
We can safely conclude that the $Au-Bu$ splitting is negligible
compared to the intrinsic width of the $\nu$ OH profile.

\begin{figure}[p]
\begin{center}
\includegraphics[width=\textwidth]{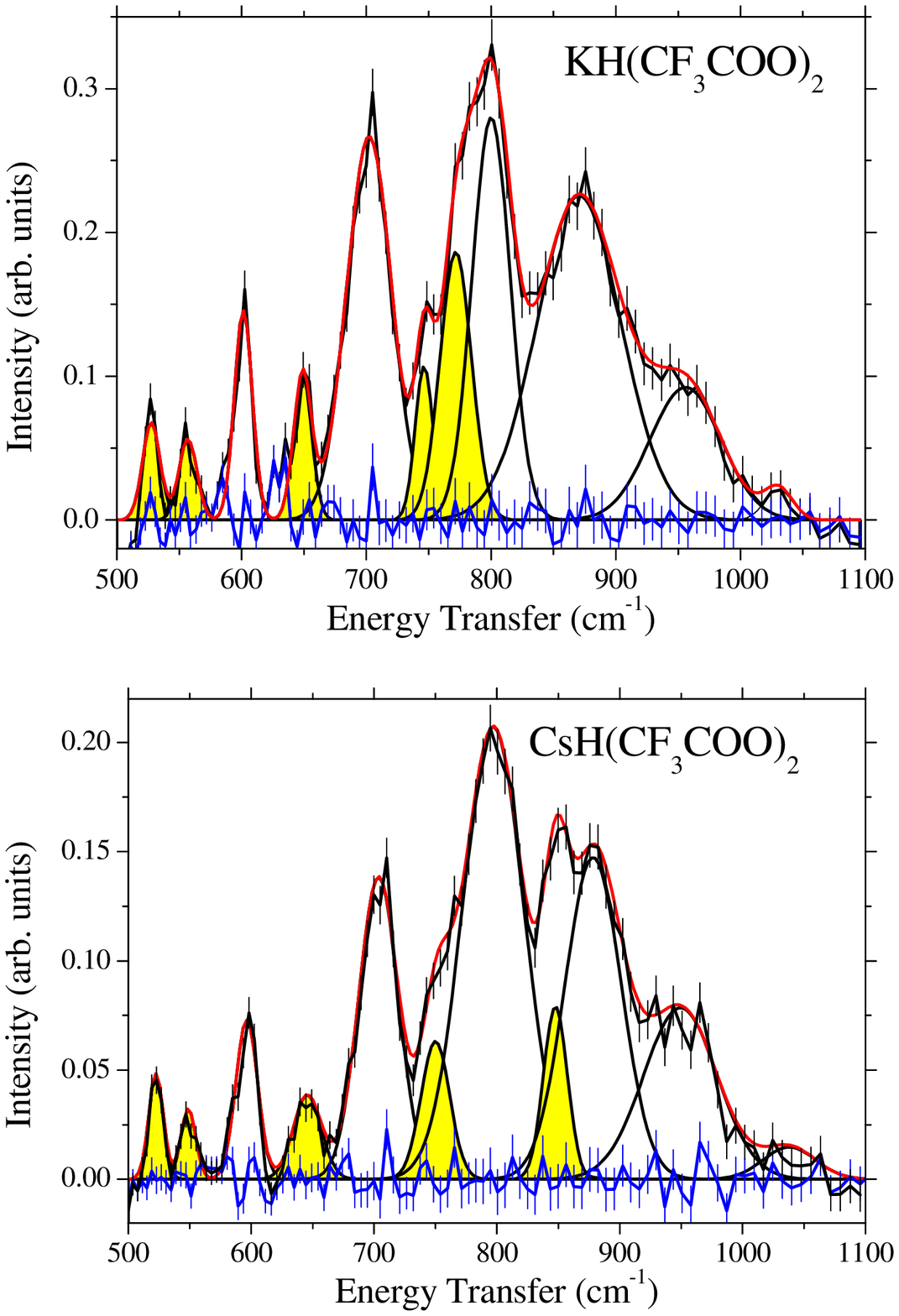}
\caption{\label{fig:05} Inelastic neutron scattering
 spectra and band decomposition into gaussian profiles
  in the OH stretching region of KH(CF$_3$COO)$_2$
  and CsH(CF$_3$COO)$_2$ at 20 K. The filled
   components are attributed to other modes (see text).
   The residual of the fit is compared to the error bars.}
\end{center}
\end{figure}

The INS spectra in the OH stretching region, from Ref. \cite{FT},
are presented in Fig.~\ref{fig:05}. The raw data have been
reanalyzed with a better background treatment. The spectra,
renormalized with respect to the amount of sample in the beam,
show very similar intensities, as anticipated for these closely
related salts. The OH stretching profiles were decomposed into
gaussian components (see Tables~\ref{tab:8} and \ref{tab:9}) and a
linear baseline was subtracted. The grey components correspond to
Raman bands previously attributed to $\delta$ CF$_3$ modes
\cite{FT}. The remaining bands, for example at $\approx 602$
(597), 704 (704), 799 (798), 871 (880) and 955 (950) \cm\ for the
K (Cs) derivative, are attributed to the OH stretching modes.

As anticipated for hydrogen bonds with equal lengths, the $\nu$ OH profiles are virtually identical.
Only in the 820-920 \cm\ range, two components at 850 and 870
\cm\ are clearly visible for the Cs derivative, while there is only one
component for the K analogue. As rather strong Raman bands were observed
at 850 and 854 \cm\ for the potassium salt \cite{FT}, we conclude that the INS component at 870 \cm\
for the K derivative is an unresolved comprise of $\delta$ CF$_3$ and $\nu$
OH modes (see Table \ref{tab:8} and \ref{tab:9}). It sounds logical that different frequencies for the
$\delta$ CF$_3$ modes arise from different conformations of the CF$_3$ groups (compare Figs. \ref{fig:03} and \ref{fig:04}).

\begin{figure}[p]
  \begin{center}
    \includegraphics[width=0.5\textwidth]{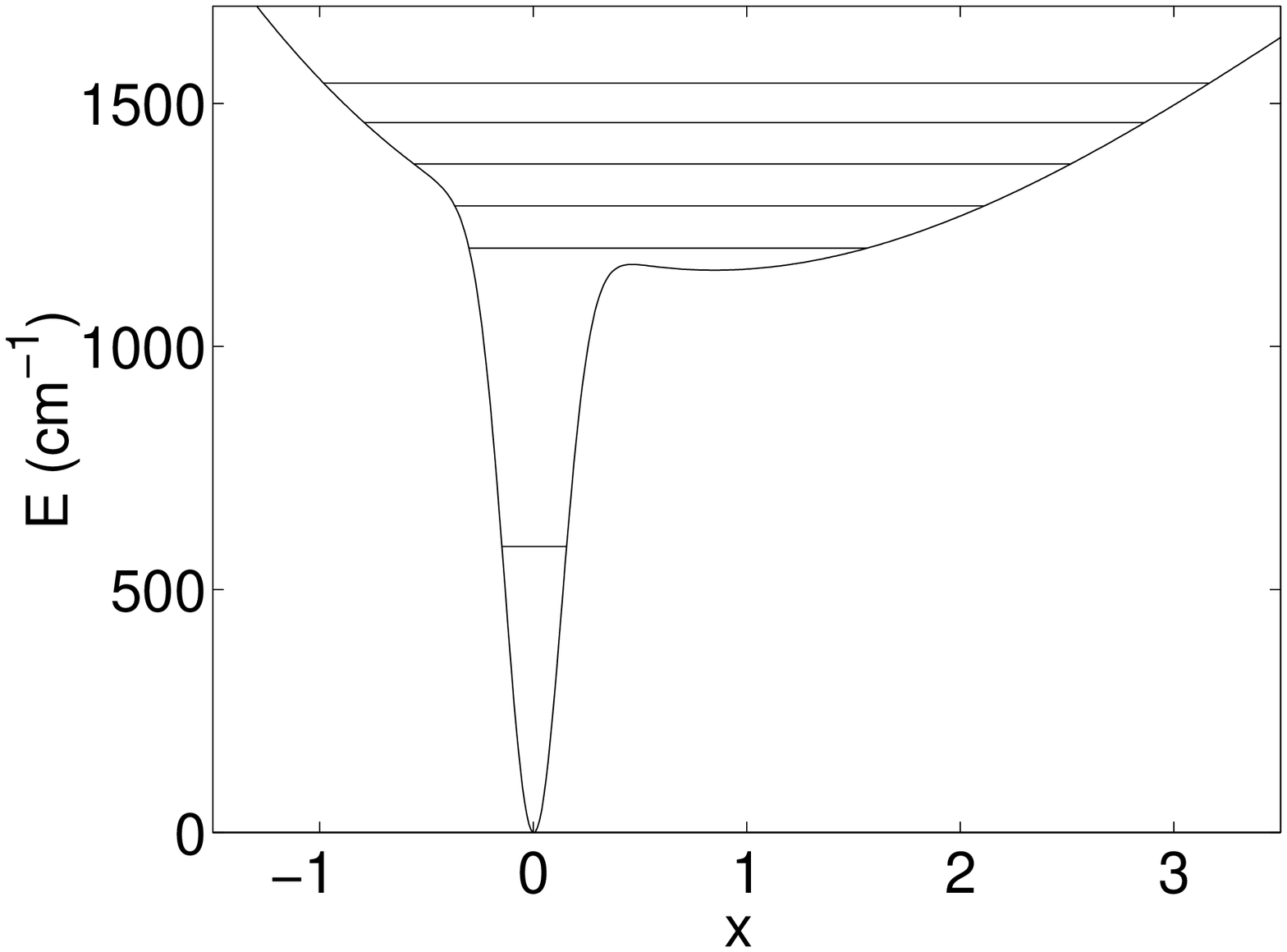}
    \includegraphics[width=0.45\textwidth]{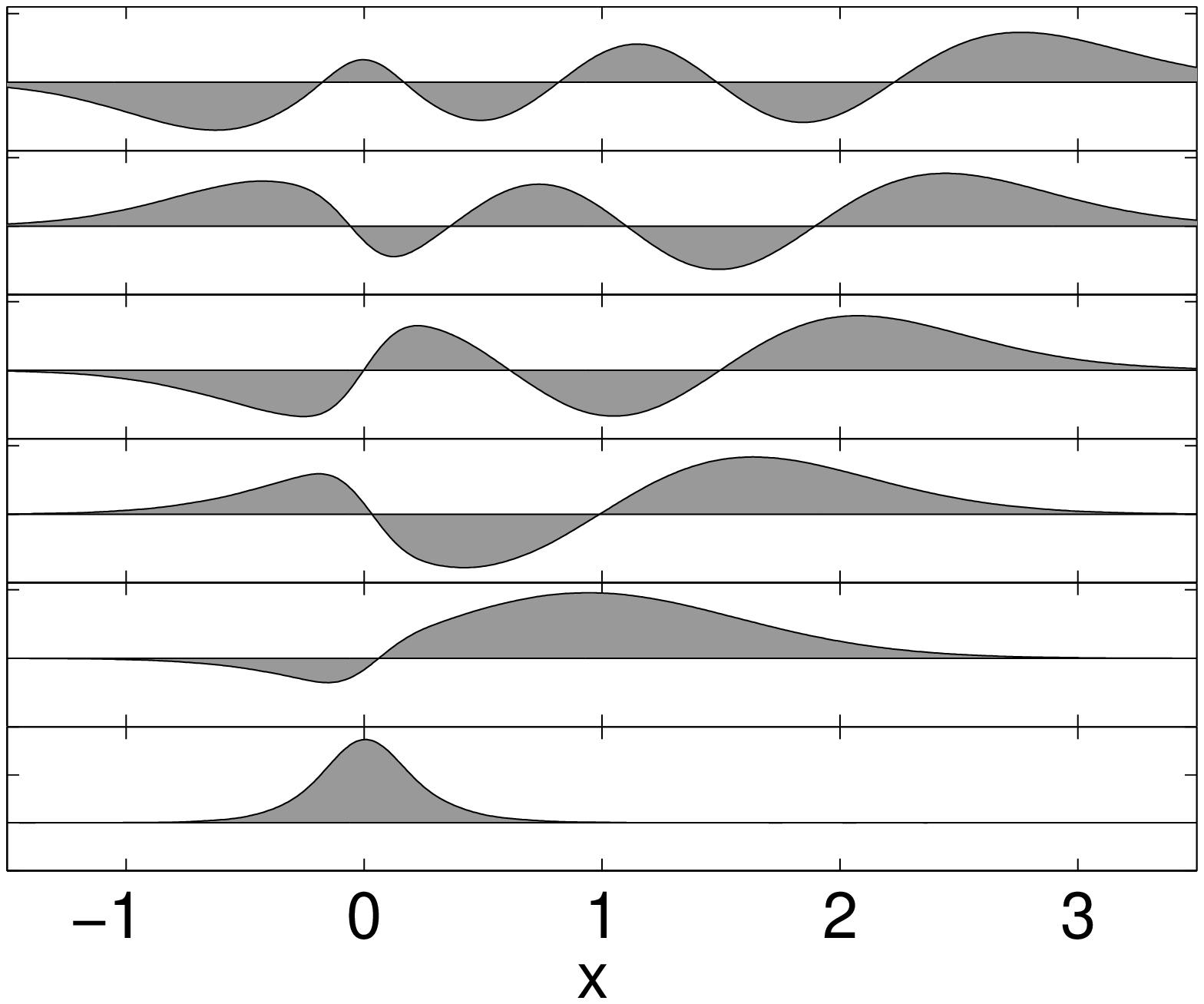}
  \caption{Left: Calculated potential and energy levels for
  KH(CF$_3$COO)$_2$. $V=-185.074\,x+ 122.598\,x^2
-10.506\,x^3-1232.04\exp(-28.149\,x^2)$, with $V$ and $x$ in
$\mathrm{cm}^{-1}$ and \AA\ units, respectively.
  Right: The corresponding wave functions, $x$ in \AA\ units.}
  \label{fig:06}
\end{center}
\end{figure}

The rather complex $\nu$ OH profiles are thus supposed to
arise primarily from the internal dynamics of the hydrogen bond. At first sight, the
slowly varying spacing of the components, namely $\approx 102$ (107), 95 (94),
72 (82), 84 (70) \cm\ in Tables \ref{tab:8} and \ref{tab:9}, suggests a Franck-Condon-like
progression due to strong coupling with O$\cdots$O modes \cite{Step,MW,SZS,HMR}. However,
the dynamical separation, analogous to the Born-Oppenheimer approximation for
electrons and nuclei, is irrelevant for strong hydrogen bonds with OH stretching
at low frequency. Therefore, by analogy with the seminal case of
potassium hydrogen maleate
\cite{FLTCP}, we assume that the effective potential function for the proton
can be decomposed into a deep and narrow well containing the ``bonding'' ground state
and a shallow quasi harmonic potential for ``non-bonding'' excited states. Then,
the relative intensities of the components
can be ascribed to a substantial shift of the minimum of the upper shallow
curve with respect to the centre of the
hydrogen bond. With this model, the best fitting potential function to the
observed spectral profiles of the K derivative is presented in Fig. \ref{fig:06}
and Table \ref{tab:10}. The fitting potential for the Cs derivative should be virtually
the same. The details of the numerical calculations are given in the Appendix (see below).

We have limited the number of adjustable parameters in the potential
function to the number of observed transitions.
Since relative intensities were also included in the model, the number of experimental
values exceeds the number of parameters. The potential function was
decomposed into a narrow gaussian well with a depth of $\approx 1000$ \cm\ and
a rather flat harmonic potential whose fundamental frequency corresponds to the mean level-spacing.
The linear term displacing the upper wave functions with respect
to the ground state allows for adjustment of relative intensities.
The small cubic term accounts for deviations from harmonicity of the upper potential.

The maximum difference between observed and calculated frequencies in Table \ref{tab:10}
is $\Delta\nu / \nu$ $\approx 2\%$. The agreement between observed and calculated intensities is
less satisfactory, with deviations as large as $\approx 30\%$. These differences
emphasize the limits of the model that ignores the complex interactions between
$\nu$ OH and $\delta$ CF$_3$ modes. Let us recall that the incoherent scattering
cross section of fluorine atoms is negligible and the coherent cross section is
rather modest ($\approx 4$ barns), compared to the incoherent cross section for
H atoms ($\approx 80$ barns) \cite{DL}. Therefore, INS intensities for the CF$_3$
modes are almost totally borrowed from the OH mode, thanks to vibrational coupling \cite{FT}.
Consequently, the estimated frequencies and intensities obtained
by simple decomposition into gaussian components presented in Tables \ref{tab:8} and \ref{tab:9} are
different from what they should be in the absence of such interactions. We did not
try to model this complex interaction. Nevertheless, we consider that coupling with CF$_3$
modes should not alter the main features of the potential function.

The potential function represented in Fig. \ref{fig:06} is totally unconventional
and we are not aware of any other reported example with similar shape. Although this
is a particular fit with a chosen analytical function, we consider that any alterative
parameterization should yield basically the same overall shape for the best fit to the
observed data.

The potential asymmetry is not in conflict with the
centrosymmetric dimers in the crystal structure. At low
temperature, only the ground state is populated. The minimum is
located at the center of the hydrogen bond and the wave function
is virtually symmetrical. At room temperature, the population of
excited states ($\lessapprox 5\%$) is too small to be
distinguished with neutron diffraction. The crystal structure
averaged over space and time remains centrosymmetric and the tiny
contribution of excited states is embedded in the temperature
factors.

In the ground state, the proton
is well localized at the centre of the hydrogen bond. The estimated vibrational mean
square amplitude of $\approx 0.02$ \AA$^2$ compares quite favorably to the
temperature factors given in Tables \ref{tab:4} and
\ref{tab:7}. In the excited states, proton transfer to one of the oxygen atoms breaks down
the symmetry of the hydrogen bond. The mean position at $\approx 1$
\AA\ from the centre is clearly incompatible with the hydrogen bond
length R$_{\mathrm{O}\cdots\mathrm{O}} \approx 2.4$ \AA\ when the proton is in the ground
state. Such a large displacement of the hydrogen atom is possible if the hydrogen
bond is broken in the excited states and the spatial extension of the
wave functions over several \AA, greater than the O$\cdots$O distance at equilibrium,
is representative of the dissociation of the salt.

\begin{figure}[p]
  \begin{center}
\includegraphics[width=\textwidth]{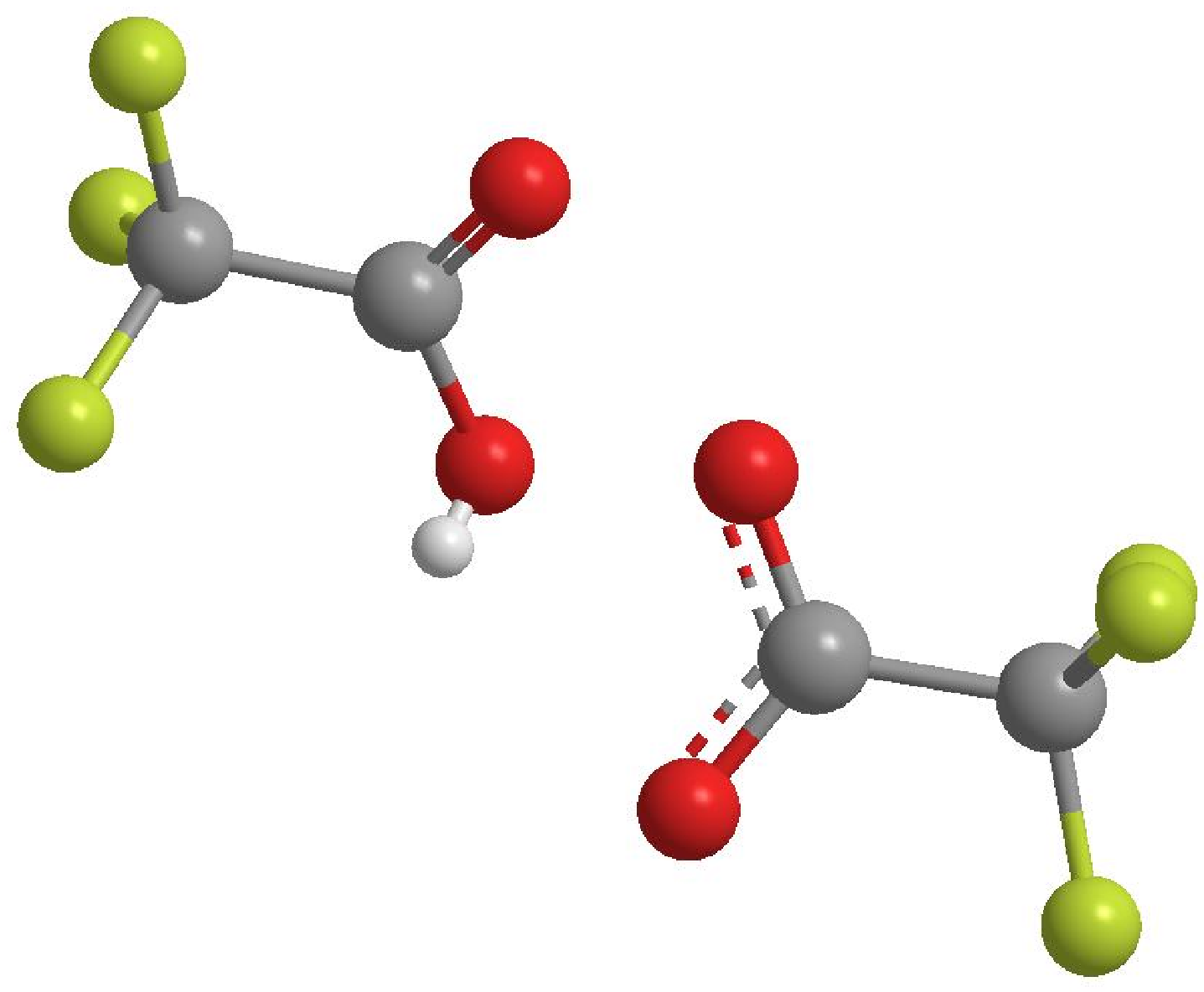}
  \caption{Artistic view of the dissociation of the hydrogen bistrifluoroacetate ion in the OH
  stretching excited state}
  \label{fig:07}
\end{center}
\end{figure}

The hydrogen bistrifluoroacetate complex in the OH excited state is tentatively represented
in Fig. \ref{fig:07}. The proton is transferred to one of the two entities. The extension of
the wave functions suggests that there is a complex combination of translation along the
hydrogen bond and rotation around the C---OH bond. At the same time, there may be some
twisting around the C---C bond to keep the proton close to the O$\cdots$O direction. These
complex dynamics are
possible because the upper level spacing of $\approx 100$ \cm\ is comparable to the frequency
of the torsional mode \cite{FT}. Proton dynamics on the one hand, internal and O$\cdots$H$\cdots$O
modes at low frequency, on the other, are on the same timescale and can interact strongly.

The asymmetry of the potential in Fig. \ref{fig:06} contrasts to the symmetric potential
for potassium hydrogen maleate \cite{FLTCP}. Presumably, the electronic structure of the
hydrogen maleate ring preserves the symmetry in the excited states. In contrast
to the naive representation of the dissociated excited state proposed in Fig 10 of Ref.
\cite{FLTCP}, the excited state should be better regarded as a quantum superposition of the
two configurations corresponding to proton transferred to one oxygen atom or the other.

The effective potential functions discussed above were determined upon the assumption
of a bare proton, $m=1$ amu, while coupling with heavy atoms were ignored. This approximation is
reasonable only for the ground state. For upper states, we should include
dynamical coupling with internal degrees of freedom of the trifluoroacetate
entities, according to the scheme in Fig. \ref{fig:07}.
Then, the effective oscillator mass should be significantly greater than 1 amu,
the effective upper potential should be steeper than shown in Fig. \ref{fig:06}
and the spatial extension of the wave functions should be diminished. The modelling of the
effective mass requires more information than available at the
present stage of investigation. However,
we suspect the overall shape should remain qualitatively the same.

Our conclusions are quite at variance from those proposed in a
recent study of the crystal structure of
4-cyano-2,2,6,6-tetramethyl-3,5-heptanedione, a new example of a
very short symmetric intramolecular hydrogen bond with
R$_{\mathrm{O}\cdots \mathrm{O}} = 2.393$ \AA\ at 100 K
\cite{BCCHK}. The assignment of the $\nu$ OH mode to an INS band
at 370 \cm\ is sustained by density functional calculations and
magic angle spinning (MAS) solid-state NMR. The authors conclude
(surprisingly) that the potential for the hydrogen bonding proton
should be of the low-barrier double minimum type, ``Although the
neutron [diffraction] data better fit a single anisotropic thermal
ellipsoid...'' Unfortunately, there is no quantitative information
on the thermal factors of the proton for comparison with other
systems and visual examination does not reveal any significant
difference compared to maleate or trifluoroacetates. Clearly,
there is not enough information to conclude on the shape of the
potential.

\section*{Conclusion}

The crystal structures of K and Cs hydrogen bistrifluoroacetates determined
with the single crystal neutron diffraction technique confirm that the very short
hydrogen bonds linking trifluoroacetate entities are symmetrical at room
temperature and at $20$ K or 14 K. There is no evidence for proton disorder or
phase transition. Any double minimum potential for the proton
can be rejected \cite{FT}.

The complex INS spectral profiles for the OH stretching modes are tentatively
decomposed into gaussian components. Comparison with Raman spectra allows
us to distinguish bands arising from the $\delta$ CF$_3$ modes in the same
frequency range. We can thus assign several OH stretching transitions and
propose a best fitting potential largely inspired by
that previously determined for potassium hydrogen maleate \cite{FLTCP}.
This potential is composed of a narrow well at the hydrogen bond centre
containing the bonding ground state and a rather flat quasiharmonic upper
part corresponding to non bonding states. The rather low energy dissociation
of this strong hydrogen bond is similar to that observed previously for
potassium hydrogen maleate. Apparently, it is a property of this type of
hydrogen bonding, rather than a consequence of the particular structure of
the maleate ring.

In contrast to potassium hydrogen maleate, symmetry breakdown occurs for
hydrogen bistrifluoroacetates upon excitation of the $\nu$ OH modes. We
conclude that the proton is transferred to one oxygen atom or the other.
Presumably, the preserved potential symmetry of the hydrogen maleate in the
excited states is a consequence of the particular structure of the ring. The
excited states should be regarded as quantum superposition of proton
transferred to the two oxygen atoms of the hydrogen bond.

We conclude that the concept of hydrogen bonding for the ground state
and hydrogen nonbonding for excited states could be of general relevance
for strong symmetric hydrogen bonds.

\section*{Appendix: Calculation of energy levels, wave functions and INS intensities}

\label{appendix1} The variational method is the most appropriate to determine
analytical potential functions fitting any given energy level
scheme. The expansion of the eigen functions with harmonic basis sets allows us to
calculate all matrix elements of interest for vibrational spectroscopy.
In this appendix we gather the formulae to
resolve the Schr\"odinger equation
for a dimensionless particle with effective mass $m^*$ experiencing
a potential $V(x)$ along the $x$ coordinate:
\begin{equation}
-\fracc{\hbar^2}{2\,m^*}\fracc{\d^2\Psi}{\d
\,x^2}+V(x)\,\Psi=E\,\Psi \label{eq:scho1}
\end{equation}

with
 \begin{equation}
 V(x)=V_p(x)+V_G(x)=\sum_{l=1}^6 v_{l}\, x^l +
 \sum_{l=1}^3a_l\exp(-b_l\,x^2)\,,
 \label{eq:fullpotential}
\end{equation}
where $V_p$ is the polynomial potential and $V_G$ the sum of three
Gaussians, to allow for a rich variety of potential forms. We also propose a straightforward method to test
the accuracy of the eigenfunctions.

In order to construct a basis set, let $\wo/(2\pi)$ be a
frequency and consider the dimensionless
variables $\xi=\alpha \,x$, with $\alpha=\sqrt{m\,\wo/\hbar}$,
$\V=V/(\hbar\,\wo)$ and $\E =E/(\hbar\,\wo)$. Then
$$\V_p=\sum_{l=1}^6 \v_l\, \xi^l,$$
with $$\v_l=v_l/(\hbar\,\wo\,\alpha^l)$$

Here, the scaled energies are such that the difference between energy
levels of the harmonic oscillator is $1$ instead of $4$ in Ref.~\cite{HRG59}.
Our constants $v_l$, divided by a factor of 2
compared to \cite{HRG59} are such that
the potential energy of the harmonic oscillator $\V=\xi^2/2$
corresponds to $\v_l=1/2\,\delta_{l,2}$ (where $\delta_{l,2}$ is the
Kronecker symbol). This choice of scaling sounds more natural and clear.
Then, $\alpha=\alpha_0\,\sqrt{\nnu_0\,m^*}$ can be expressed in
\AA$^{-1}$ units with $\alpha_0=0.17273$ and
$\nnu_0=\wo/(2\,\pi\,c)$ in \cm\ units.

In the new variables Eq.~(\ref{eq:scho1}) reads:
\begin{equation}
\hh\,\psi=-\fracc{1}{2}\fracc{\d^2\psi}{\d
\,\xi^2}+\V\,\psi=\E\,\psi\;, \label{eq:scho2}
\end{equation}
with $\psi =\sqrt{\alpha}\,\Psi$ normalized with respect to $\xi$.

For a harmonic oscillator the normalized solutions of Eq.~\ref{eq:scho2}
are:
\begin{equation}
u_n(\xi)=(2^n\,n!\sqrt{\pi})^{-1/2}\,H_n(\xi)\exp
(-\xi^2/2)\quad\;\quad n=0,1,\dots \,,\label{eq:harmonic}
\end{equation}
$\{H_n\}$ being the Hermite polynomials. The functions $\{u_n\}$
form a suitable orthonormal basis set for localized solutions
$\psi$. For numerical calculations the dimension $N$ of the basis
has to be finite. We found $N=60$ is suitable to calculate the 10
lowest energy levels with good accuracy (see below). The matrix
elements of $\hh$ in Eq.~\ref{eq:scho2} are:
$$\hh_{n,m}=\langle n |\hh|m\rangle =\int_{-\infty}^{\infty}
u_n(\xi)\hh\,u_m(\xi)\,\d\,\xi\,.$$

The matrix elements used to calculate the eigenstates for a given
oscillator mass and a sixth order polynomial potential or a
Gaussian potential  can be found in Refs.~\cite{HRG59} and
\cite{CS63}, respectively. As these papers may be nowadays
difficult to obtain, we present the algorithms  in condensed form,
with an alternative definition of the scaling factors for the
polynomial form.

For the polynomial potential $V=V_p$ we obtain:
\begin{equation}
\begin{array}{lcl}
\hh^p_{n,n}&=&(n+\fracc{1}{2})\,(\fracc{1}{2}+\v_2)+
\fracc{3}{4}\,(2\,n^2+2\,n+1)\,\v_4\\
& + & \fracc{1}{8}\,(20\,n^3+30\,n^2+40\,n+15)\,\v_6\;; \nonumber\\
\hh^p_{n,n-1}&=&\fracc{1}{2}\sqrt{2\,n}\,
\Big(\v_1+\fracc{3}{2}\,n\,\v_3+\fracc{5}{4}\,(2\,n^2+1)\,\v_5\Big)\;;\nonumber\\
\hh^p_{n,n-2}&=&\fracc{1}{2}\sqrt{n\,(n-1)}\,\Big(-\fracc{1}{2}+
\v_2+(2\,n-1)\,\v_4 \\
& + & \fracc{15}{4} \,(n^2-n-1)\,\v_6\Big)\;;
\nonumber \\
\hh^p_{n,n-3}&=& \fracc{1}{2}
\sqrt{\fracc{n\,(n-1)\,(n-2)}{2}}\,\Big(\v_3+\fracc{5}{2}\,(n-1)\,\v_5\Big)
\;;\nonumber\\
 \hh^p_{n,n-4}&=&
\fracc{1}{4}\,\sqrt{n\,(n-1)\,(n-2)\,(n-3)}\,\,
\Big(\v_4+\fracc{3}{4}\,(2\,n-3)\,\v_6\Big)
\;;\nonumber\\
\hh^p_{n,n-5}&=&\fracc{1}{4}\,
\sqrt{\fracc{n\,(n-1)\,(n-2)\,(n-3)\,(n-4)}{2}}\,\,\v_5
\;;\nonumber\\
\hh^p_{n,n-6}&=&\fracc{1}{8}\,
\sqrt{n\,(n-1)\,(n-2)\,(n-3)\,(n-4)\,(n-5)}\,\,\v_6 \;.
\label{eq:HH}
\end{array}
\end{equation}
The terms independent on $\v_l$ arise from the kinetic energy. All
other matrix elements are zero except for the symmetric ones
$\hh^p_{n-l,n}=\hh^p_{n,n-l}$, $l=1,\dots,6$.

For a Gaussian potential term $W=a\,\exp(-b\,x^2)$, the scaled
potential is $\tW=\a\,\exp(-\b\,\xi^2)$, with $\a=a/(\hbar\,\wo)$
and $\b=b/\alpha^2$. Matrix elements $\WW_{n,m}$ in the subset
$\{u_n\}_{n=0}^{N-1}$ are calculated according to
Ref.~\cite{CS63}:
  construct a
$(N,2N-1)$ auxiliary matrix $\GG$ with an iterative procedure.
The first row is calculated as:
\begin{equation}
\GG_{0,m}=\sqrt{\frac{m!} {2^m(1+\b)} }\,\frac{(-\theta)^{m/2}} {(m/2)!}
\quad \mbox{for m\,(even)=0,\dots, 2N-1} \nonumber \;,
\end{equation}
with $\theta=\b/(1+\b)$. The second row is calculated as:
\begin{equation}
 \GG_{1,m}=\sqrt{m}\,\GG_{0,m-1}+\sqrt{m+1}\,\GG_{0,m+1}
 \quad ;\mbox{for m (odd)=1,\dots,2N-2} \nonumber \;
\end{equation}
Each successive row of index $n$ ($2 \leqq n \leqq N-1$) depends
on the two previous ones as:
\begin{eqnarray*}
\GG_{n,m} & = & 1/\sqrt{n}\,(\sqrt{m}\,\GG_{n-1,m-1}+\sqrt{m+1}\,\GG_{n-1,m+1}
       -\sqrt{n-1}\,\GG_{n-2,m});\\
m & = & n,n+2,n+4,\dots,2N-n-1
\end{eqnarray*}
All elements not explicitly assigned are set to zero. Let us
redefine $\GG$ as the $(N,N)$ square matrix corresponding to its
first $N$ columns. The procedure above has led to an upper
triangular matrix. The elements of the lower triangle are obtained
by symmetry $\GG(n,m)=\GG(m,n)$, for all $n,m$ such that $m<n$.
The $(N,N)$ matrix corresponding to $\tW$ is
  $\WW=\a\,\GG(\b)$.

Therefore, the matrix $\hh$ corresponding to the full potential
$V$ in Eq.~\ref{eq:fullpotential} is:
$$ \hh=\hh^p +\sum_{l=1}^3 \a_l\,\GG(\b_l)$$
If $\{\E_n\}_{n=0}^{N-1}$ are eigenvalues in increasing order, and
$\CC$ is the $(N,N)$ matrix whose column $n$ is the normalized
eigenvector corresponding to $\E_n$, the eigenfunctions are
$$\psi_n(\xi)=\sum_{m=0}^{N-1}\,\CC_{n,m} \,u_m(\xi).$$ The
eigenvalues and eigenfunctions in physical units are
$$E_n=\hbar\,\wo\,\E_n$$
and $$\Psi_n(x)=\sqrt{\alpha}\sum_{m=0}^{N-1}\,\CC_{n,m}
\,u_m(\alpha\,x).$$
 The eigenfunctions $\Psi_n$ are analytical functions
(although with numerically calculated coefficients)
composed of a polynomial of order $N-1$ multiplied by a Gaussian.
Therefore, derivatives are easily obtained and the accuracy
of the solutions can be checked by substitution in
Eq.~(\ref{eq:scho1}). As a rule of thumb, the last coefficients of
each series $\{\CC_{n,m}\}_{m=0}^{N-1}$ (say, the last 10 for
$N=60$) have to be very
small with respect to the largest of the others. This is generally
achieved for the first few eigenfunctions to be compared with observation.

The parameter $\wo$ largely determines whether the truncated expansions
of $\{ \Psi_n\}$ are good approximations. As a rule of thumb the
exponential in Eq.~\ref{eq:harmonic} should be small (say $\sim
\mathrm{e}^{-2}$) at the estimated limits for the the particle
position. This leads to $\wo \approx 16\,\hbar/(m^*\,\Delta x^2)$,
where $\Delta x$ is the width of the classically allowed region,
or $\nnu_0 \approx 16/(m^* \alpha_0^2 \Delta x^2)$ with $\nnu_0$
and $\Delta x$ in $\mathrm{cm}^{-1}$ and \AA\ units, respectively.
For a single minimum this is obtained if $\hbar \,\wo$ is close to
the first observed transition. In this case, basis sets
limited to sizes $N\approx 40$ are convenient. For potentials
composed of a narrow well and a shallow upper part, like those for
strong symmetric hydrogen bonds under consideration in this paper,
it is necessary to increase the size to $N = 60$. The accuracy for
the 10 lower eigenvalues is far beyond experimental errors and the
accuracy of the eigenvectors is better than $1\%$. Further
increment of $N$ is unnecessary as numerical errors
become larger the higher powers of $x$.

The INS intensity for a transition $|0\rangle \to |n\rangle$ at
energy $E_n$ is proportional to the scattering function
$$S(Q,E)=|\langle 0 |\exp(-iQx)|n \rangle |^2\delta(E-E_n).$$
For a spectrometer like TFXA \cite{TFXA}, energy and momentum transfer,
$E$ and $Q$, respectively, are correlated as
$$E \approx \frac {h^2Q^2}{16.759m*},$$
with $E$ and $Q$ in \cm\ and \A\ units, respectively.

Therefore, for a given set of parameters $\{v_l\}$, $\{a_l\}$,
$\{b_l\}$ and $\wo$, we are able to obtain the frequencies and
intensities of the first few transitions with good accuracy. Using
standard numerical methods it is possible to determine the values
of the parameters that best fit the observed frequencies and
intensities.


\begin{table}[p]
\caption{\label{tab:1} Neutron single crystal diffraction data
and structure refinement for potassium and cesium hydrogen bistrifluoroacetates.
 $\lambda$ = 0.8305 \AA. Space groups monoclinic $I 2/a$ for potassium and $A2/a$ for cesium. Both
 with Z = 4. The criterion for observed reflections was I $>$ 3$\sigma $(I).
  The variance for the last digit is given in parentheses.}
\ \\
\begin{center}
\begin{tabular}{llllllll}
\hline
& \multicolumn{3}{l}{KH(CF$_3$COO)$_2$} & \ & \multicolumn{3}{l}{CsH(CF$_3$COO)$_2$} \\
 \cline{2-4} \cline{6-8}
 & 20K & \ \ & 298 K & & 14 K & \ & 298 K\\
 \cline{2-2} \cline{4-4} \cline{6-6} \cline{8-8}
a ({\AA}) & 8.68(1) & & 8.78(1) & & 13.44(1) & & 13.623(8) \\
b ({\AA}) & 10.023(9) & & 10.18(1) & & 4.942(9) & & 5.033(3)\\
c ({\AA}) & 9.146(9) & & 9.28(1)  & & 14.35(1) & & 14.741(6) \\
\textit{$\beta $ }(\r{ }) & 100.36(8) & & 99.96(9)  & & 112.88(9) & & 112.46(9) \\
$V$ ({\AA}$^{3})$ & 782.6 & & 817.0 & & 878.4 & & 934.0\\
$D_{x}$ (Mg m$^{-3})$ & 2.259 & & 2.164  & & 2.722 & & 2.559 \\
 & & & & & & & \\
Measured reflections & 1979 & & 2164  & & 3476 & & 3897 \\
Independent reflections & 1769 & & 1472  & & 1977 & & 2096 \\
and observed reflections & 1503 & & 1142  & & 1557 & & 1062 \\
$R_{int}$ & 0.037 & & 0.038  & & 0.048 & & 0.064 \\
& & & & & & & \\
Refinement on  & F & & F & & F & & F \\
R-factor & 0.040 & & 0.045  & & 0.0399 & & 0.043 \\
Weighted R-factor & 0.042 & & 0.029  & & 0.0345 & & 0.040 \\
Goodness of fit & 1.070 & & 1.088  & & 1.049 & & 1.079 \\
Number of reflections & 1503 & & 1142 & & 1557 & & 1062 \\
Number of parameters & 76 & & 76  & & 76 & & 76 \\
used in refinement & & & & & & & \\
Extinction coefficient & 22.7(7) & & 17.2(8)  & & 10.0(3) & & 11.2(9) \\

\hline
\end{tabular}
\end{center}
\end{table}

\begin{table}[p]
\caption{\label{tab:2} Atomic positions and isotropic
temperature factors for KH(CF$_3$COO)$_2$ at 20 K
(first lines) and 298 K (second lines).
The variance for the last digit is given in parentheses.}

\ \\

\begin{center}
\begin{tabular}{lllll}
\hline
 Atom & $x/a$ & $y/b$ & $z/c$ & U(iso)(\AA$^2$) \\
\hline
K(1) & 0.2500     & 0.47035(13)& 1.0000 & 0.0022 \\
     & 0.2500     & 0.4691(2)  & 1.0000 & 0.0306 \\
C(1) & 0.08108(6) & 0.65992(5) & 0.61071(5) & 0.0020 \\
     & 0.0771(1)  & 0.66238(9) & 0.61201(9) & 0.0306 \\
C(2) & -0.04133(6) & 0.66059(5) & 0.71503(6) & 0.0020 \\
     & -0.04356(9) & 0.66241(8) & 0.71512(8) & 0.0260 \\
O(1) & -0.02359(7) & 0.58604(6) & 0.82262(7) & 0.0040 \\
     & -0.02706(13) & 0.58871(11) & 0.81966(11)& 0.0333 \\
O(2) & -0.15042(7) & 0.74496(6) & 0.67180(7) & 0.0045 \\
     & -0.15011(14) & 0.74556(14) & 0.67428(14)& 0.0426 \\
F(1) & 0.15118(7)  & 0.78039(7) & 0.61351(7) & 0.0050 \\
     & 0.14415(18) & 0.78021(15) & 0.61321(18)& 0.0519 \\
F(2) & 0.19203(7)  & 0.56895(7) & 0.65128(7) & 0.0046 \\
     & 0.18745(15) & 0.57426(17) & 0.65321(17)& 0.0490 \\
F(3) & 0.01440(8)  & 0.63619(7) & 0.47010(7) & 0.0046 \\
     & 0.01325(19) & 0.63720(16) & 0.47512(13)& 0.0474 \\
H(1) & -0.2500     & 0.7500     & 0.7500     & 0.0166 \\
     & -0.2500     & 0.7500     & 0.7500     & 0.0574 \\
\hline
\end{tabular}
\end{center}
\end{table}

\begin{table}[p]
\caption{\label{tab:3} Interatomic distances in \AA\ units
 and angles in degrees in KH(CF$_3$COO)$_2$ at 20 K
 (first lines) and 298 K (second lines).
 The variance for the last digit is given in parentheses. }

\ \\

\begin{center}
\begin{tabular}{llll}
\hline
C(1)--C(2) & 1.5504(17) & C(1)--F(1) & 1.3503(15) \\
           & 1.5455(19) &            & 1.335(2) \\
C(1)--F(2) & 1.3296(16) & C(1)--F(3) & 1.3332(17) \\
           & 1.327(2)   &            & 1.322(2) \\
C(2)--O(1) & 1.2230(15) & C(2)--O(2) & 1.2779(16) \\
           & 1.2148(18) &            & 1.270(2) \\
O(2)--H(1) & 1.2179(13) & C(2)--C(1) - F(1) & 109.9(1) \\
           & 1.2159(17) &                   & 110.08(14) \\
C(2)--C(1) - F(2) & 111.9(1)   & C(2)--C(1) --F(3) & 111.57(11) \\
                  & 111.51(14) &                   & 111.87(13)\\
F(1)--C(1) - F(2) & 107.86(12) & F(1)--C(1)--F(3)  & 107.08(11) \\
                  & 107.78(17) &                   & 107.18(16) \\
F(2)--C(1) - F(3) & 108.29(11) & C(1)--C(2)--O(1)  & 119.4(1) \\
                  & 108.24(16) &                   & 119.61(12) \\
C(1)--C(2)--O(2)  & 111.4(1)   & O(1)--C(2)--O(2)  & 129.21(9) \\
                  & 111.36(13) &                   & 129.03(12) \\
C(2)--O(2)--H(1)  & 114.2(1)   & O(2)--H(1)--O(2)  & 180.0(1) \\
                  & 114.87(13) &                   & 180.0(1) \\
\hline
\end{tabular}
\end{center}
\end{table}

\begin{table}[p]
\caption{\label{tab:4} Thermal parameters in \AA$^2$ units
 for KH(CF$_3$COO)$_2$ at 20 K (first lines) and 298 K
 (second lines). The variance for the last digit is
 given in parentheses. }

\ \\

\begin{center}
\begin{tabular}{lllllll}

\hline
 Atom & U$_{11}$ & U$_{22}$ & U$_{33}$ & U$_{23}$ & U$_{13}$ & U$_{12}$ \\
\hline
K(1) & 0.0027(4)   & 0.0018(4) & 0.0020(4) & 0.0000 & 0.0003(3) & 0.0000 \\
     & 0.035(1)    & 0.0263(8) & 0.0325(9) & 0.0000 & 0.0122(7) & 0.0000 \\
C(1) & 0.00117(19) & 0.00268(19) & 0.0023(2) & 0.00024(13) & 0.00039(14) & 0.00003(13) \\
     & 0.0297(4)   & 0.0329(4)   & 0.0299(3) & 0.0029(3)   & 0.0078(3)   & -0.0022(3) \\
C(2) & 0.00157(19) & 0.00238(19) & 0.0023(2) & 0.00055(14) & 0.00060(14) & 0.00064(13)\\
     & 0.0258(3)   & 0.0272(3)   & 0.0249(3) & 0.0045(3)   & 0.0043(2)   & 0.0005(3) \\
O(1) & 0.0038(2) & 0.0048(2) & 0.0035(2) & 0.00241(17) & 0.00120(17) & 0.00108(15)\\
     & 0.0360(5) & 0.0341(4) & 0.0307(4) & 0.0115(4)   & 0.0088(3)   & 0.0042(4) \\
O(2) & 0.0033(2) & 0.0060(2) & 0.0047(2) & 0.00268(17) & 0.00194(17) & 0.00312(16)\\
     & 0.0376(5) & 0.0507(7) & 0.0418(5) & 0.0194(5)   & 0.0129(4)   & 0.0180(4) \\
F(1) & 0.0049(2) & 0.0048(2) & 0.0054(2) & 0.00042(18) & 0.00072(18) & -0.00222(17)\\
     & 0.0523(8) & 0.0457(7) & 0.0597(8) & 0.0074(6)   & 0.0150(6)   & -0.0195(6) \\
F(2) & 0.0027(2) & 0.0054(2) & 0.0057(2) & 0.00058(18) & 0.00076(18) & 0.00219(17)\\
     & 0.0353(6) & 0.0579(8) & 0.0557(7) & 0.0057(6)   & 0.0135(5)   & 0.0141(5) \\
F(3) & 0.0054(2) & 0.0057(2) & 0.0025(2) & -0.00028(17) & 0.00020(17) & -0.00006(17)\\
     & 0.0608(8) & 0.0535(7) & 0.0285(5) & -0.0032(5)  & 0.0094(5)    & 0.0005(6) \\
H(1) & 0.0177(7) & 0.0158(8) & 0.0160(7) & 0.0034(6)    & 0.0023(6)   & 0.0029(6)\\
     & 0.0539(17) & 0.0577(18) & 0.0582(17) & 0.0199(15) & 0.0034(13) & 0.0152(14) \\
\hline
\end{tabular}
\end{center}
\end{table}

\begin{table}[p]
\caption{\label{tab:5} Atomic positions and isotropic
temperature factors for CsH(CF$_3$COO)$_2$ at 14 K
(first lines) and 298 K (second lines).
The variance for the last digit is given in parentheses.}

\ \\

\begin{center}
\begin{tabular}{lllll}
\hline
 Atom & $x/a$ & $y/b$ & $z/c$ & U(iso)(\AA$^2$) \\
\hline
    Cs(1)    &     -0.2500      &  0.21541(18) &  0.0000     & 0.0013 \\
             &     -0.2500      &  0.2170(4)   &  0.0000     & 0.0377 \\
    F(1)     &     -0.01883(5)  &  0.61975(15) &  0.15253(5) & 0.0059 \\
             &     -0.01376(14) &  0.5971(4)   &  0.15586(14)& 0.0652 \\
    F(2)     &      0.10983(5)  &  0.39738(15) &  0.26553(5) & 0.0069 \\
             &      0.11435(18) &  0.3806(5)   &  0.26110(12)& 0.0761 \\
    F(3)     &      0.14684(5)  &  0.72718(14) &  0.18559(5) & 0.0063 \\
             &      0.14651(17) &  0.7048(4)   &  0.18308(15)& 0.0722 \\
    O(1)     &      0.16871(5)  &  0.28677(13) &  0.08444(5) & 0.0050 \\
             &      0.16517(9)  &  0.2837(3)   &  0.0796(1)  & 0.0445 \\
    O(2)     &     -0.00501(5)  &  0.18925(14) &  0.05235(5) & 0.0045 \\
             &     -0.00336(8)  &  0.1803(3)   &  0.0543(1)  & 0.0442 \\
    C(1)     &      0.07995(4)  &  0.52050(11) &  0.17554(4) & 0.0031 \\
             &      0.08238(8)  &  0.5005(2)   &  0.17453(7) & 0.0401 \\
    C(2)     &      0.08431(4)  &  0.31424(11) &  0.09597(4) & 0.0027 \\
             &      0.08415(6)  &  0.30433(18) &  0.09498(7) & 0.0312 \\
    H(1)     &      0.0000      &  0.0000      &  0.0000     & 0.0166 \\
             &      0.0000      &  0.0000      &  0.0000     & 0.0592 \\

\hline
\end{tabular}
\end{center}
\end{table}

\begin{table}[p]
\caption{\label{tab:6} Interatomic distances in \AA\ units
 and angles in degrees in CsH(CF$_3$COO)$_2$ at 20 K
 (first lines) and 298 K (second lines).
 The variance for the last digit is given in parentheses. }

\ \\

\begin{center}
\begin{tabular}{llll}
\hline
   C(1)--C(2)    &     1.549(3)  &  C(1)--F(1)    &     1.331(3)  \\
                 &     1.5403(14)&                &     1.3236(19) \\
   C(1)--F(2)    &     1.340(3)  &  C(1)--F(3)    &     1.331(2)  \\
                 &     1.326(2)  &                &     1.324(2)  \\
   C(2)--O(1)    &     1.216(3)  &  C(2)--O(2)    &     1.277(3)  \\
                 &     1.2132(13)&                &     1.2746(13)\\
   O(2)--H(1)    &     1.218(2)  &  C(2)--C(1)--F(1)       &    112.3(2) \\
                 &     1.2221(12)&                         &    112.3(1) \\
  C(2)--C(1)--F(2)   &    109.20(17) &  C(2)--C(1)--F(3)   &    111.56(15) \\
                     &    110.24(13) &                     &    111.63(11) \\
  F(1)--C(1)--F(2)   &    107.6(2)   &   F(1)--C(1)--F(3)  &    108.10(12) \\
                     &    107.61(15) &                     &    107.41(16) \\
  F(2)--C(1)--F(3)   &    107.9(2)   &  C(1)--C(2)--O(1)   &    118.59(17) \\
                     &    107.44(16) &                     &    118.5(1) \\
  C(1)--C(2)--O(2)   &    112.5(2)   &   O(1)--C(2)--O(2)  &    128.92(15) \\
                     &    112.86(8)  &                     &    128.62(11) \\
  C(2)--O(2)--H(1)   &    114.03(19) &   O(2)--H(1)--O(2)  &    180.0(1) \\
                     &    114.18(9)  &                     &    180.0(1) \\

\hline
\end{tabular}
\end{center}
\end{table}

\begin{table}[p]
\caption{\label{tab:7} Thermal parameters in \AA$^2$ units
 for CsH(CF$_3$COO)$_2$ at 14 K (first lines) and 298 K
 (second lines). The variance for the last digit is
 given in parentheses. }

\ \\

\begin{center}
\begin{tabular}{lllllll}

\hline
 Atom & U$_{11}$ & U$_{22}$ & U$_{33}$ & U$_{23}$ & U$_{13}$ & U$_{12}$ \\
\hline
Cs(1)   &     0.0009(3)  & 0.0015(3)  & 0.0019(3)   & 0.0000      & 0.0009(2)   &  0.0000 \\
        &     0.0319(6)  & 0.0359(7)  & 0.0526(9)   & 0.0000      & 0.0244(6)   & 0.0000 \\
F(1)    &     0.0048(2)  & 0.0069(3)  & 0.0061(2)   &-0.00185(18) & 0.00239(17) &  0.00167(19) \\
        &     0.0614(8)  & 0.069(1)   & 0.068(1)    &-0.0250(8)   & 0.0275(7)   & 0.0110(8) \\
F(2)    &     0.0095(2)  & 0.0082(3)  & 0.0030(2)   & 0.00077(19) & 0.00246(17) &  0.0006(2) \\
        &     0.0959(13) & 0.0907(14) & 0.0369(7)   & 0.0065(8)   & 0.0199(8)   & 0.0014(12) \\
F(3)    &     0.0074(2)  & 0.0047(3)  & 0.0068(3)   &-0.00216(19) & 0.00288(18) & -0.00294(18) \\
        &     0.0837(12) & 0.0581(9)  & 0.0777(12)  &-0.0291(9)   & 0.0342(9)   & -0.0324(9) \\
O(1)    &     0.0038(2)  & 0.0055(2)  & 0.0071(2)   &-0.00185(18) & 0.00373(16) & -0.00097(18) \\
        &     0.0344(5)  & 0.0458(6)  & 0.0592(7)   &-0.0061(6)   & 0.0245(5)   & -0.0035(5) \\
O(2)    &     0.0029(2)  & 0.0039(2)  & 0.0068(2)   &-0.00313(18) & 0.00219(16) & -0.00094(17) \\
        &     0.0305(4)  & 0.0407(6)  & 0.0634(7)   &-0.0202(6)   & 0.0202(5)   & -0.0076(4) \\
C(1)    &     0.00369(17)& 0.0030(2)  & 0.00295(18) &-0.00084(14) & 0.00151(13) &  0.00004(15) \\
        &     0.0448(5)  & 0.0407(5)  & 0.0346(4)   &-0.0067(4)   & 0.0148(3)   & -0.0051(4) \\
C(2)    &     0.00267(18)& 0.00290(19)& 0.00315(18) &-0.00089(15) & 0.00183(13) & -0.00025(14) \\
        &     0.0285(3)  & 0.0284(3)  & 0.0370(4)   &-0.0012(3)   & 0.0130(3)   & -0.0011(3) \\
H(1)    &     0.0131(6)  & 0.0176(8)  & 0.0185(7)   & 0.0006(7)   & 0.0053(5)   & -0.0011(6) \\
        &     0.0403(13) & 0.0498(16) & 0.088(2)    &-0.0059(17)  & 0.0247(14)  & -0.0011(12) \\

\hline
\end{tabular}
\end{center}
\end{table}

\begin{table}[p]
\begin{center}
\caption{Gaussian decomposition of the INS spectrum of
KH(CF$_3$COO)$_2$ in the OH stretching region (see Fig. \ref{fig:05}).}
\label{tab:8}

\ \\

\begin{tabular}{|c|c|c|c|c|}
 \hline
Peak & Width & Gravity center & Area  & Assignment \\
 \mbox{}& \cm &\cm& \%&\mbox{}\\
\hline
  1 & 12 & 527 & 1.9 &$\delta$ CF$_3$\\
  2 & 36 & 560 & 3.4 &  $\delta$ CF$_3$\\
  3 & 18 & 602 & 4.4 & OH stretching\\
  4 & 36 & 648 & 5.1 &  $\delta$ CF$_3$\\
  5 & 41 & 704 & 17.9 & OH stretching\\
  6 & 22 & 750 & 5.0 & $\delta$ CF$_3$\\
  7 & 20 & 774 & 5.0 & $\delta$ CF$_3$\\
  8 & 38 & 799 & 18.1 &OH stretching \\
  9 & 78 & 871 & 27.6 &OH stretching, $\delta$ CF$_3$\\
  10& 73 & 955 & 10.9 &OH stretching\\
  11 &18 & 1033 & 0.8 &OH stretching \\
   \hline
\end{tabular}
\end{center}
\end{table}

\begin{table}[p]
\begin{center}
\caption{Gaussian decomposition of the INS spectrum of
CsH(CF$_3$COO)$_2$ in the OH stretching region (see Fig. \ref{fig:05}).}
\label{tab:9}

\ \\

\begin{tabular}{|c|c|c|c|c|}
 \hline
Peak & Width  & Gravity center & Area  & Assignment \\
\mbox{}& \cm &\cm& \%&\mbox{}\\
 \hline
  1 & 12 & 523 & 1.8 &$\delta$ CF$_3$\\
  2 & 19 & 550 & 1.5 &  $\delta$ CF$_3$\\
  3 & 20 & 597 & 3.9 & OH stretching\\
  4 & 26 & 647 & 2.6 &  $\delta$ CF$_3$\\
  5 & 38 & 704 & 14.2 & OH stretching\\
  6 & 27 & 751 & 4.7 & $\delta$ CF$_3$\\
  7 & 57 & 798 & 30.6 & OH stretching\\
  8 & 23 & 849 & 5.1  & $\delta$ CF$_3$\\
  9 & 53 & 880 & 20.1 &OH stretching\\
  10 &68 & 950 & 13.4 &OH stretching\\
  11 &55 & 1043 & 2.0 &OH stretching \\
   \hline
\end{tabular}
\end{center}
\end{table}

\begin{table}[p]
\begin{center}
\caption{Observed and calculated INS OH stretching frequencies and
intensities for KH(CF$_3$COO)$_2$. The calculated ones correspond
to the potential $V=-185.074\,x+ 122.598\,x^2
-10.506\,x^3-1232.04\exp(-28.149\,x^2)$, $V$ and $x$ in \cm and
\AA, respectively. For the TFXA spectrometer used in Ref. \cite{FT},
energy and momentum transfer are correlated as $\nnu = Q^2/(16.759 m)$
with $\nnu$, $Q$ and $m$ in \cm, \A\ and amu units, respectively. au: arbitrary units.}
\label{tab:10}

\ \\

\begin{tabular}{|c|c|c|c|c|c|c|}
  \hline
  \mbox{}
 & \multicolumn{3}{|c|}{Observations}& \multicolumn{3}{|c|}{Calculations} \\
\cline{2-7}

 Transitions& freq. &  Q &  int.  & freq. &  Q &  int. \\
\mbox{}& (cm$^{-1}$) &  (\AA$^{-1}$) &  (au)  & (cm$^{-1}$) &  (\AA$^{-1}$) &  (au) \\
\hline
  0 $\rightarrow$ 1 & 601 & 6.0  & 0.25  & 614  &  6.1  &  0.38  \\
  0 $\rightarrow$ 2 & 703 & 6.5  & 0.96  & 701  &  6.5  &  0.93 \\
  0 $\rightarrow$ 3 & 798 & 6.9  & 1.00  & 787  &  6.9  &  1.00 \\
  0 $\rightarrow$ 4 & 870 &  7.2 & 0.76  & 872  &  7.2  &  0.59 \\
  0 $\rightarrow$ 5 & 956 &  7.6 &  0.58 & 953  &  7.5  &   0.37\\
  \hline
\end{tabular}
\end{center}
\end{table}

\clearpage

\end{document}